\documentclass[aip,apl,reprint]{revtex4-1}
\usepackage{graphicx}
\usepackage{bm}
\usepackage{pgfplots}
\usepackage{float}
\usepackage{color}
\usepackage{ulem}

\begin{document}
\title{\textcolor{black}{Weak ferromagnetism and short range polar order in NaMnF$_3$ thin films}}
\author{Amit KC}
\affiliation{Department of Physics, West Virginia University, Morgantown, WV 26506, USA}
\affiliation{Department of Physics, University of California, Santa Cruz, CA 95064, USA}
\author{Pavel Borisov}
\affiliation{Department of Physics, West Virginia University, Morgantown, WV 26506, USA}
\author{Vladimir V. Shvartsman}
\affiliation{Institute for Materials Science, University of Duisburg-Essen, Universit\"{a}tsstra{\ss}e 15, 45141 Essen, Germany}
\author{David Lederman}
\affiliation{Department of Physics, West Virginia University, Morgantown, WV 26506, USA}
\affiliation{Department of Physics, University of California, Santa Cruz, CA 95064, USA}

\begin{abstract}
The orthorhombically distorted perovskite NaMnF$_{3}$ has been predicted to become ferroelectric if \textcolor{black}{an $a=c$ distortion of the bulk \textit{Pnma} structure} is imposed.  In order to test this prediction, NaMnF$_{3}$ thin films were grown on SrTiO$_{3}$ (001) single crystal substrates via molecular beam epitaxy.
The best films were smooth and single phase with four different twin domains. 
In-plane magnetization measurements revealed the presence of 
antiferromagnetic ordering with weak ferromagnetism below the N\'eel temperature $T_N=66$~K. For the 
dielectric studies, NaMnF$_{3}$ films were grown on a 30 nm SrRuO$_{3}$ (001) layer
used as a bottom electrode grown via pulsed laser deposition. 
The complex permittivity as a function of frequency 
indicated a strong Debye-like relaxation contribution characterized by a distribution of relaxation times. A power-law divergence of the characteristic relaxation time revealed an order-disorder phase transition at 8~K. The slow relaxation dynamics indicated the formation of super-dipoles \textcolor{black}{(superparaelectric moments)} that extend over several unit cells, similar to polar nanoregions of relaxor ferroelectrics. 
\end{abstract}
\date{\today}
\maketitle

%\clearpage

%\tableofcontents

%\section{Introduction}

There is much interest in multiferroic (MF) materials that show coexistence of two or more long-range orders such as ferroelectricity, (anti-) ferromagnetism and/or ferroelasticity.  Magnetoelectric (ME) coupling can allow for the control of magnetization $\mathbf{M}$ (electric polarization $\mathbf{P}$) using an electric field $\mathbf{E}$ (a magnetic field $\mathbf{H}$), which can be used for potential applications in data storage, sensors and spintronic devices.
\cite{eerenstein2006multiferroic, fiebig2005revival, Cherepov, cheong2007multiferroics, Hayes} 
As a result, there has been much effort expended towards synthesizing MF insulating materials with strong ME coupling at room temperature.\cite{zhao2006electrical, park2007size,Yang, Hwang} 

Many MF materials are oxides, but multiferroicity can also be found in fluorides. For example, 
in the orthorhombic Ba$M$F$_4$ family   
the $M=$ Co and Ni compounds in bulk form are ferroelectric (FE) and antiferromagnetic.\cite{scott2011multiferroic}
While in oxide perovskites ABO$_3$ the simultaneous presence of 
FE and magnetic orderings is limited by conflicting requirements for the 
$d^{n}$ electronic configuration of the transition metal  in the B-site,\cite{cohen1992origin} ferroelectricity in BaMnF$_4$ is mainly due to geometric (topological) reasons.\cite{ravez2000ferroelectricity} 
Recent experiments
showed that BaCoF$_4$ thin films are weakly ferromagnetic at 
low temperatures due to strain.\cite{borisov2016}  

\begin{figure}
\centering
\includegraphics[]{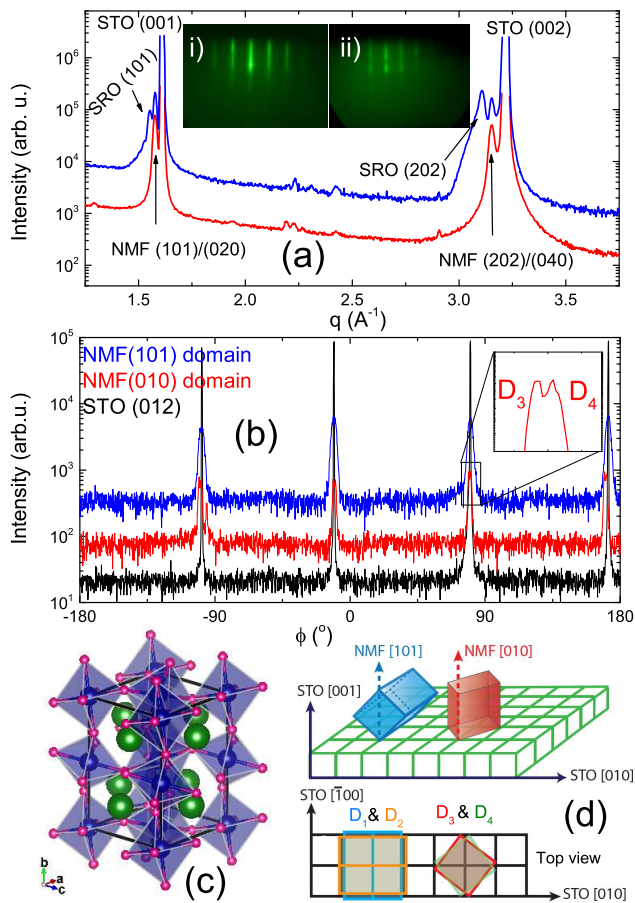}
%\caption{(a) NMF orthorhombic unit cell with MnF$_6$ octahedra. Green, blue and pink spheres are Na, Mn and F atoms, respectively. (b) Typical $\theta-2\theta$ scan of 50~nm NMF film on (100) STO. Inset shows 150~nm (red curve) NMF grown on 35~nm SRO and 50~nm NMF on bare STO (black curve) data. Red and black dots indicate NMF and STO peaks, respectively. (c) $\phi$-scans of STO (210) (black curve), NMF (222) with (010) film orientation (red curve) and NMF (222) with (101) film orientation (blue curve). Inset shows $\phi$-scan of the (010) film orientation indicating a twinning of the NMF (222) reflection. (d) Rocking curve of NMF on STO (black) and NMF on SRO/STO with $\text{FWHM}=1.32^{\circ}$ and $2.23^{\circ}$. (e) Low angle x-ray reflectivity data (open circles) for 50 nm NMF film and the fit to the data (red curve). Bottom inset shows a $2\times 2\ \mu\text{m}^2$ AFM image and top inset shows RHEED patterns of the NMF/STO (right) and the NMF/SRO films (left). (f) Schematic of the epitaxial orientation of 50 nm NMF on STO. (g) Top view schematic of NMF film for (101) and (010) film orientation with additional twin domains as seen in $\phi$-scans.}
\caption{(a) $\theta-2\theta$ XRD scan of 50~nm NMF film on STO(001) and 150~nm NMF grown on 30~nm SRO on STO(001). Insets show RHEED patterns of the NMF/STO (i) and NMF/SRO (ii) films. (b) \textcolor{black}{NMF/STO $\phi$-scans corresponding to (012) reflections (black curve) of the substrate and $(222)_\textrm{o}$ reflections of the NMF(010) (red) and NMF(101) (blue) domains with $\chi$ (the angle between the scattering vector and surface normal) $=$ $26.57^{\circ}$, $63.50^{\circ}$, and $26.50^{\circ}$, respectively. Inset shows a closeup of the NMF(010) domain indicating twinning of the $(222)_\textrm{o}$ planes. (c) NMF orthorhombic unit cell with MnF$_6$ octahedra. Green, blue and pink spheres are Na, Mn and F ions, respectively. (d) Schematic of the epitaxial orientation of NMF/STO with top views of  NMF(101) and NMF(010) domains determined from $\phi$-scans.}}
\label{fig1}
\end{figure}

NaMnF$_{3}$ (NMF) is another fluoride compound that is possibly MF with ME coupling.
%compound  belongs to  \textcolor{black}{ABF$_3$ perovskites, where the A-site is usually  K or Na and 
%the B-site is a transition metal. 
The  NMF crystal structure
is described by the orthorhombic space group \textit{Pnma}. At room temperature, the lattice constants are $a_\textrm{o}= 5.751$~\AA, $b_\textrm{o}= 8.008$~\AA, and $c_\textrm{o}= 5.548$~\AA\  (Fig.~\ref{fig1}a).\cite{ratuszna1989structure} In the pseudo-cubic unit cell, the corresponding lattice parameters are $a_{pc}= c_{pc}= \sqrt{a_\textrm{o}^{2} + c_\textrm{o}^{2}} /2  = 3.995$~\AA\ and $b_{pc} = b_\textrm{o}/2 =4.004$~\AA.  NMF 
 exhibits \textit{G}-type antiferromagnetism of the Mn magnetic moments 
 centered within tilted MnF$_6$ octahedra, with additional weak ferromagnetic 
 canting.\cite{ratuszna1989structure, pickart1964magnetic, 
 shane1967antiferromagnetic} \textcolor{black}{The N\'eel temperature has
  been reported to be $T_N=66$~K. \cite{teaney1962antiferromagnetic, shane1967antiferromagnetic, pickart1964magnetic, PhysRevB.51.12337, Everett, Du}} 
	%NMF does not have a 
 %phase transition related to a crystal symmetry transformation, but it does have a non-linear thermal expansion coefficient.\cite{katrusiak1992phase} 
 Recent computational work predicted a ferroelectric instability in NMF 
originating from Na displacements (the A-site perovskite cation in 
the ABF$_3$ structure). 
Calculations assuming that a cubic substrate forces the $a$ and $c$ lattice 
parameters to be equal to each other have shown that the soft 
polar mode $B_{2u}$ freezes, resulting in a transformation to the polar $Pna2_{1}$ space group which causes a polarization of $P=6$~$\mu$C/cm$^2$ along the long axis $b$. 
Calculations also indicate that $P$  
 can be enhanced by negative or positive strains (12~$\mu$Ccm$^2$ at $+5$\% 
 strain 9 $\mu$C/cm$^2$ at $-5$\% strain).\cite{garcia2014geometric, 
 PhysRevLett.116.117202} A weak ferromagnetic phase and an amplification of ME coupling  are also expected 
in the strained films, with a ME 
 response comparable to Cr$_2$O$_{3}$.\cite{PhysRevLett.116.117202}
 Therefore, NMF is an attractive ME material because its ferroelectricity can be modulated by strain and is not subject to the $d^{n}$ rule because there is no displacement of the Mn ion from the center of the MnF$_6$ octahedra.  

Here we report on the 
%\sout{\textcolor{black}{epitaxial}} 
growth of NMF thin films on cubic SrTiO$_{3}$ (STO) substrates and on SrRuO$_3$ (SRO) layers, \textcolor{black}{their weak ferromagnetism (FM) below the $T_N$, and their short range dipolar order}. 
%The magnetization measurements suggest that the films are weakly_{}
%ferromagnetic below 66~K, confirming theoretical predictions
%and an earlier experimental observation of antiferromagnetic resonance.\cite{shane1967antiferromagnetic} 
%The dielectric properties were measured in 150~nm thick NaMnF$_{3}$ films grown on a metallic SrRuO$_3$(SRO) (100) buffer layer. 
%The dielectric measurements indicate an order-disorder type phase transition at 
%approximately 8~K. 
%In displacive ferroelectrics, the soft phonon mode 
%propagates through the crystal, driving the transition, but in the case of the 
%order-disorder system the soft mode diffuses and there is only large amplitude 
%hoping motion.\cite{kittel2005introduction} This results in the formation of polar 
%microscopic regions which can be viewed as microregions having their own 
%ferroelectric polarizations but with slightly different Curie temperatures. 
%\cite{smolenskii1970physical} 
%It should also be noted that on these kind of 
%pervoskites, the size and degree of such ordering is govern by B-site \cite{setter1980role}.  
%\section{Experimental details}
NMF thin films approximately 50~nm thick were grown 
on pre-polished single crystal (001) STO substrates ($a=3.905$~\AA)
%obtained pre-polished from MTI Corporation) 
by molecular beam epitaxy (MBE) \textcolor{black}{by co-depositing NaF and MnF$_2$} in an ultra-high vacuum chamber with growth pressure  $P\approx 5.0 \times 10^{-9}$~Torr. 
Atomically flat surface and single termination of STO 
substrates was achieved by the combination of two thermal annealing steps and 
a de-ionized (DI) water treatment.\cite{connell2012preparation}
%At the first step, the STO substrates were annealed at 1000~$^{\circ}$C for 1 hour in air, which was followed by DI water treatment for 30~s via agitation. At the final step, the substrates were annealed at 1000 $^{\circ}$C for 1 hour in air to achieve an atomically smooth surface. 
%Atomic force microscopy (AFM) images of the substrates revealed terraces with 
%root mean square roughness ranging from 0.8 to 2.5~\AA. 
%\sout{The NMF thin films were grown via co-deposition of NaF (99.99\%) and MnF$_2$ (99.99\%) using commercial Knudsen cells.The fluxes of NaF ($\approx 0.027$~\AA/s)and MnF$_2$ ($\approx 0.043$~\AA/s) were measured using a quartz crystal monitor placed at the sample growth position. The growth was performed at substrate temperatures $T_s$ ranging from 200~$^{\circ}$C to 450~$^{\circ}$C in 50 $^{\circ}$C steps, while the quality of the substrate and film surfaces was monitored \textit{in-situ} using reflection high energy electron diffraction (RHEED).} 
\textcolor{black} {For additional details of the growth process, see supplementary material.} 
The crystallography of the substrate and film surfaces was monitored \textit{in-situ} using reflection high energy electron diffraction (RHEED). Further crystallographic characterization was carried out \textit{ex-situ} using x-ray diffraction (XRD) and x-ray reflectivity (XRR) techniques 
using a rotating anode source with a graphite bent crystal monochromator optimized for Cu K$_\alpha$ radiation. 
%Film thickness was confirmed by x-ray reflectivity (XRR) measurements. 
%using the same XRD source but with the sample mounted on a two-circle 
%Huber 
%goniometer. 
The surface morphology was studied by atomic force microscopy (AFM). Magnetic measurements were performed using a superconducting quantum interference device (SQUID) magnetometer with the substrate signal subtracted by performing identical measurements on a blank substrate.
% from Quantum Design. 

Dielectric measurements of NMF films were performed by first depositing a 30~nm SRO film on the STO substrate as a bottom electrical contact via pulsed laser deposition (PLD) in a separate vacuum chamber.  The SRO was grown at a substrate temperature of 600~$^{\circ}$C in 100 mTorr O$_2$ partial pressure. A 150 nm NMF thin film was then deposited on the SRO using the conditions described above.  
%\textcolor[rgb]{1,0,0}{\sout{Indium was used as top electrode.}}
The dielectric constant was measured using a
%Signal Recovery model 7265 
lock-in amplifier in the frequency range of 100 Hz to 100 kHz from 10 K to room temperature using a cryostat 
%Cryomech model GB15 
%helium closed cycle refrigerator 
and an ac voltage amplitude of $V_{ac}=100$~mV. Similar results were obtained for $V_{ac}=10$~mV.
% or 10 mV. 
%The responses for these two different amplitudes were almost identical so all the data discussed here were obtained with 100 mV ac amplitude. 

%\section{Experimental Results and Discussion}

%The growth was optimized by varying the substrate 
%temperature $T_{s}$ while keeping other parameters constant. 
%The films were grown at $T_{s}$ ranging from 200 to 450~$^{\circ}$C at 50~$^{\circ}$C steps. 
The best film quality was achieved at $T_{s}$= 250 and 300~$^{\circ}$C as determined from XRD and RHEED. Figure~\ref{fig1}a shows the $\theta-2\theta$ XRD scan for a film grown at $T_s= 300$~$^{\circ}$C. 
%The inset shows XRD data of a 150~nm NMF on 35~nm SRO (NMF/SRO) (red curve) along with 50~nm NMF on bare (100) STO (NMF/STO) (black curve); red dots represent NMF peak. 
The out-of-plane lattice parameters were 3.989~\AA~and 3.982~\AA, and the rocking curve full width at half maximum (FWHM) values were 1.32$^{\circ}$ and 2.23$^{\circ}$ for NMF/STO and NMF/SRO films, respectively (see supplementary material). The positions of the SRO peaks were consistent with data in the literature.\cite{Kim2001} 
%The observed peaks, corresponding to the pseudo-cubic film (100) reflection, can be assigned either to (101) or (010) orthorhombic orientations. 
%The rocking curve had a full width at half maximum (FWHM) of 1.32$^{\circ}$ with a Lorentzian fit (Fig.~\ref{fig1}d). 
XRR data (supplementary material) were fitted using GenX software for the NMF/STO film\cite{Bjorck:aj5091} to obtain a NMF thickness of 51.76~nm and a NMF surface surface roughness of 0.4~nm. Streaky RHEED patterns (Fig.~\ref{fig1}a inset i) confirmed that the NMF/STO film was smooth and crystalline while the NMF/SRO sample also had streaky patterns with a somewhat fuzzy streaky background (Fig.~\ref{fig1}a inset ii). %Samples grown at other temperatures had higher roughness parameters 
%as verified by RHEED and AFM, 
%and/or were polycrystalline.
% as indicated by formation of spots and ring-like patterns in the corresponding RHEED patterns. 
%On the other hand, films grown at lower temperatures  ($T_s=$ 250-300 $^{\circ}$C) were smoother as indicated by formation of streaky pattern. Top inset shows AFM image of NMF film grown with the same total thickness (50 nm).

\textcolor{black}{
%Using the \textit{Pnma} setting (as discussed above)  for NMF, the XRD $\phi$-scans and reciprocal space maps (RSMs) indicates the film have grown textured as NMF(010)/STO(001) and NMF(101)/STO(001) (the corresponding lattice parameters are virtually identical, see Fig.~\ref{fig1} and supplementary material Fig. S4). 
Using the \textit{Pnma} NMF structure, reflections associated with $b_\textrm{o}$-axis pointing out-of-plane [NMF(010)/STO(001)] as well as in-plane [NMF(101)/STO(001)] domains were identified via XRD $\phi$-scans and reciprocal space maps (RSMs) (the corresponding (010) and (101) lattice parameters are virtually identical; see supplementary material Fig.~S4 for RSM data).
These results indicate the absence of $a_\textrm{o}$ = $c_\textrm{o}$ 
distortion in the NMF/STO film for both of these domains. The measured lattice parameters for NMF(010)/STO(001) and NMF(101)/STO(001) domains were $a_\textrm{o}= 5.76$~\AA, $b_\textrm{o}= 7.98$~\AA, and $c_\textrm{o}= 5.57$~\AA\ and $a_\textrm{o}= 5.86$~\AA, $b_\textrm{o}= 8.04$~\AA, and $c_\textrm{o}= 5.45$~\AA\, respectively (see Table S1 in the supplementary material).} \textcolor{black}{Figure~\ref{fig1}b shows the $\phi$-scans of the $(222)_\textrm{o}$ reflection of the NMF(101) and NMF(010) domains along with (012) reflections of the substrate. In both NMF domains, four peaks separated by $90^{\circ}$ were observed, meaning that two NMF phases with orthogonal $b_\textrm{o}$ axes were present. For NMF(010) domains, the orthorhombic $a_\textrm{o}$ and $c_\textrm{o}$ axes $\parallel$ STO [110]. For NMF(101) domains, the $b_\textrm{o}$ axis commensurately aligns $\parallel$ STO [010] and [$\bar{1}$00] (labeled D1 and D2 in Fig.~\ref{fig1}d). 
%Therefore, the NMF(010) and NMF(101) domains grew on STO square lattices corresponding to $a\sqrt{2}\times a\sqrt{2}$ and $2a\times2a$, respectively ($a=3.905$~\AA).
%Using \textit{Pnma} setting for NMF reflections associated with $b_\textrm{o}$-axis pointing out-of-plane as well as in-plane domains were identified (the corresponding lattice parameters are virtually identical). This result ruling out $a_\textrm{o}$ = $c_\textrm{o}$ distortion in the film for both of these domains.For NMF/STO, only reflections associated with orthorhombic phase could be identified ruling out $a_\textrm{o}$ = $c_\textrm{o}$ distortion in the film. The films had both $\{010\}_\textrm{o}$ and $\{101\}_\textrm{o}$ out-of-plane orthorhombic lattice reflections, i.e., with the $b_\textrm{o}$-axis pointing out-of-plane or in-plane (the corresponding lattice parameters are virtually identical). Using \textit{Pnma} setting for NMF as discussed above, XRD $\phi$-scans and reciprocal space maps (RSMs) determined the orientations and lattice parameters (see supplementary material Fig. S4 and table SI).}  Figure~\ref{fig1}b shows the $\phi$-scan for the (210) reflection of STO together with the $(222)_\textrm{o}$ reflection of NMF. 
% \textcolor{black}{Refer to Fig. S5 table for computed lattice parameters for both of these orientations.} 
For the NMF(010) domains, each $(222)_\textrm{o}$ reflection was additionally comprised of two peaks separated by $1.4^{\circ}$ (Fig.~\ref{fig1}b, inset). Therefore, each NMF(010) domain was actually comprised of two in-plane twin domains  rotated by $1.4^{\circ}$ about the surface normal with respect to each other (labeled D$_3$ and D$_4$ in Fig.~\ref{fig1}b inset and ~\ref{fig1}d top view). 
%No rotated in-plane domains were observed for the (101) oriented  phase.
%The orthorhombic lattice parameters can be described by epitaxy relations $a_\textrm{o}^2+c_\textrm{o}^2\approx (2a_{\text{STO}})^2$ and $b_\textrm{o} \approx 2a_{\text{STO}}$ for \textit{b} in-plane and \textit{b} out-of-plane orientations, respectively. 
Similar epitaxies of orthorhombically distorted perovskite oxide films on STO have been reported elsewhere.\cite{Kim2001,marti2008crystal, daumont2009epitaxial, scola2011microstructure}}

\textcolor{black}{For the NMF/SRO sample, epitaxy could not be confirmed by $\phi$-scans or RSM, despite the streaky RHEED pattern in Fig.~\ref{fig1}a. This indicates that the bulk of NMF was highly disordered, with small domains. Because of the similarity between STO and SRO, however, it is reasonable to assume that the strain in the NMF/SRO sample was not too different from the NMF/STO sample. }

%For NMF/SRO, in-plane XRD confirmed the presence of both (101) and (010) phases similar to NMF/STO, but no rotated in-plane domains for the (010) oriented phase were observed (for additional XRD and AFM structural data, see supplementary material). 
%The ratio between these two (101) and (010) orthogonal phases was found to be 1.1$\pm$0.5 and 2.6$\pm$0.7 for NMF/STO and NMF/SRO respectively. Computational work has also found that (101) phase is supposedly less polar than the (010) phase.\cite{Garcia}

\begin{figure}
\begin{center}
\includegraphics[]{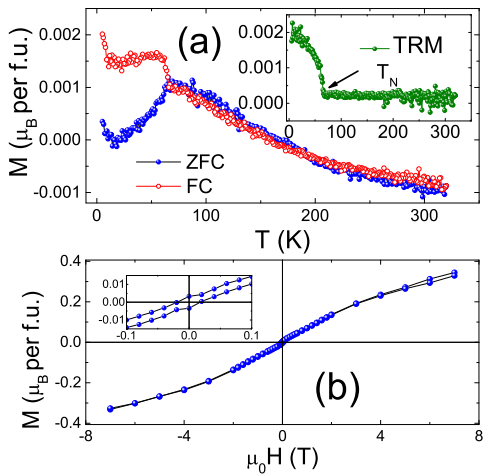}
\end{center}
\caption{Magnetic behavior of a 50 nm NMF/STO thin film. (a) FC and ZFC magnetic moment $m$ as a function of temperature $T$. Inset shows TRM data. The negative FC and ZFC background is likely a result of an imperfect subtraction of the substrate signal. (b) In-plane magnetization $M$ as a function of the applied magnetic field $\mu_0 H$ at 5~K. Inset shows close-up view of the low-field region.}
\label{fig2}
\end{figure}
%The magnetic properties were studied on 50~nm NMF films. 
%grown at $T_s= 300$~$^{\circ}$C.  
The temperature-dependent magnetization measurements on NMF/STO were carried out by first cooling the 
samples from room temperature to 5~K in $H=0$, and then measuring the magnetic moment $m$ as a 
function of $T$ in $H=1$~kOe applied in-plane along the STO [001] 
direction while warming up to 320~K (zero-field cooled, ZFC). 
Then the measurements were continued while cooling in the same $H$ down to 5 K (field 
cooled, FC).  Subsequently, the thermoremanent magnetization (TRM) was measured while warming in $H=0$. 
%All magnetization data were corrected for the substrate magnetic response, measured separately under identical conditions, and normalized to the film volume. 
ZFC and FC data in Fig.~\ref{fig2}a show a Curie-Weiss-like increase with decreasing temperature until the ZFC data peak at $T_N\approx 66$~K.  On the other hand, the FC $m$ decreases 
sharply below $T_N\approx 66$~K, \textcolor{black}{in good agreement with theory\cite{garcia2014geometric} and previous experimental work.\cite{teaney1962antiferromagnetic, shane1967antiferromagnetic, pickart1964magnetic, PhysRevB.51.12337, Everett}} This split between ZFC and FC magnetizations below 66~K is consistent with weak FM below 
$T_N$, \textcolor{black}{in agreement with previous studies of NMF
nanoparticles and nanoplates}.\cite{Everett,Du}  
%A second transition previously reported in bulk NMF at $T\approx 145$~K was not observed.\cite{teaney1962antiferromagnetic} 
The TRM data have a Brillouin-like increase below $T_N$, confirming the weak ferromagnetic nature of the transition. 
%sBecause bulk MnF$_2$ has collinear antiferromagnet order below $T_N\approx 67$~K ,\cite{leighton1999competing} and no indication of MnF$_2$ crystalline order was found in the XRD scans, we can exclude this possibility. 
Fig~\ref{fig2}b shows the in-plane magnetization hysteresis loop $M (H)$ measured at 5 K with the diamagnetic response from the substrate subtracted. The $M(H)$ behavior is consistent with weak FM, including an S-shaped loop with a lack of saturation in magnetic fields of up to 7 T and a remanent magnetization $M (H = 0)\approx 0.003\ \mu_B/\text{f.u.}$ in the low field region (Fig.~\ref{fig2}b, inset).  
\begin{figure}
\begin{center}
\includegraphics[]{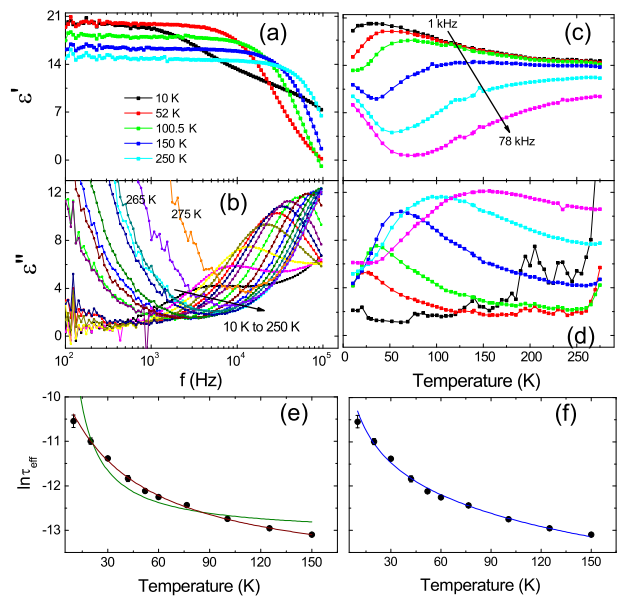}
\end{center}
\caption{Real (a) and imaginary (b) parts of the dielectric constant $\varepsilon$ of the 150~nm NMF/SRO film as a function of frequency $f$ at different temperatures $T$. 
%The arrow points to increasing temperatures from 10 K to 250 K in (b). 
Real (c) and imaginary (d) parts of $\varepsilon (T)$ at $f=$ 1, 5, 10, 25, 50 and 78~kHz. Arrow in (c) points to the direction of increasing $f$. (e) Arrhenius law (olive curve), Vogel-Fulcher law (wine curve), and (f) power law (blue curve) fits of the temperature dependence of the relaxation time.}
\label{fig3}
\end{figure} 

Figure~\ref{fig3}c and d shows the dielectric permittivity $\varepsilon$ as a function of $T$ for the 150 nm thick NMF/SRO film at frequencies from 1 to 78 kHz. The response at 0.1 kHz almost overlaps with that at 1 kHz so it is not shown. At large frequencies, the increase of the real part of the dielectric permittivity 
$\varepsilon^\prime(T)$ at low $T$ is superimposed with a broad maximum whose position shifts to lower $T$ with decreasing frequency $f$. The imaginary part of dielectric permittivity, $\varepsilon^{\prime\prime}(T)$, also shows a peak that shifts to lower $T$ and becomes narrower with decreasing $f$. This behavior implies a strong relaxation of the dielectric permittivity and resembles properties of relaxor 
ferroelectrics,\cite{bokov2006recent} but the frequency dependence of the position of the maximum of $\varepsilon^\prime(T)$, $T_{m}$, 
is much stronger than 
is typically observed in relaxors.\cite{Samara2003} 
In our case, $T_m$ shifts more than 200 K in only two frequency decades. 

To gain more insight in relaxation dynamics, we analyzed $f$ dependence of $\varepsilon$. 
%Figure~\ref{fig3} a and b summarizes frequency dependence of the real and imaginary parts of complex dielectric permittivity measured at different temperatures. 
At low $f$, $\varepsilon^{\prime\prime}(f)$ follows a $1/f$ dependence which is due to a 
non-zero dc conductivity (Fig.~\ref{fig3}b). 
Also, a typical relaxation maximum is observed for $f> 3$~kHz, whose position shifts to 
lower $f$ as $T$ decreases. The $\varepsilon^\prime(f)$ behavior shows a step-like increase at 
frequencies where $\varepsilon^{\prime\prime}(f)$ is a maximum. As $T$ decreases below 52 K, another relaxation peak seems to appear at a higher frequency. 
However, this peak was not within our experimental window down to the lowest $T$, and only the low-$f$ wing is visible. 

Assuming a Debye-like dielectric relaxation, the broad peaks (steps) in the 
$\varepsilon^{\prime\prime}(f)$ ($\varepsilon^{\prime}(f)$) data indicate the existence of a distribution of relaxation times. Using the Cole-Cole dispersion law,\cite{kremer2003broadband}
% the permeattivity can be described by
\begin{equation}
\varepsilon (f) = \varepsilon_{\infty} + \frac{\Delta\varepsilon}{1 + (i2 \pi f\tau_{\text{eff}})^{\beta}},
\end{equation}
where $\tau_{\text{eff}}$ is an effective characteristic relaxation time, 
$\varepsilon_{\infty}$ is the dielectric permittivity at very high frequencies, 
$\Delta\varepsilon$ is the relaxation strength, and $\beta$ ($0<\beta\le 1$) indicates the width of the relaxation time distribution ($\beta= 1$ corresponds to a Debye relaxation with a single relaxation time). The values of $\tau_{\text{eff}}(T)$, extracted from the position of the maximum of 
$\varepsilon^{\prime\prime}(f)$, $\tau_{\text{eff}}= (2\pi f_{m})^{-1}$, are shown in Figs.~\ref{fig3}e-f. $\tau_{\text{eff}}(T)$ cannot be satisfactory fitted \textcolor{black}{either} by an Arrhenius law, 
\begin{equation}
\tau_{\text{eff}}(T) = \tau_{0}\exp(-E_{a}/kT), 
\label{eq:Arr}
\end{equation}
\textcolor{black}{or} by a Vogel-Fulcher law applicable to typical relaxor FE,
\begin{equation}
\tau_{\text{eff}}(T) = \tau_{0}\exp[-E_{a}/k(T-T_{f})],
\label{eq:VF}
\end{equation}
 where $k$ is Boltzmann's constant and $E_a$ is an activation energy, as seen 
 in Fig.~\ref{fig3}e.  In the case of Eq.~\ref{eq:Arr}, the quality of the fit is low, while the fit to Eq.~\ref{eq:VF} results in an unphysical negative value of the freezing temperature, $T_{f}$.
On the other hand, $\tau_{\text{eff}}(T)$ is described well by the power law equation 
\begin{equation}
\tau_{\text{eff}}(T)=\tau_{0}(T/T_{0}-1)^{\nu}.
\end{equation}
The best fit (Fig.~\ref{fig3}f) yields $\nu = -0.90\pm 0.07$, $\tau_0=(2.8\pm1.5)\times 10^{-5}$ s, and a transition temperature $T_{0}= 8$~K. The value of $\nu$ is close to $\nu=-1$ that describes the divergence of the relaxation time for an order-disorder phase transition in the classical 
mean-field approximation.\cite{lines2004principles} The actual transition temperature could be much lower than 8~K because only the behavior for $T>T_0$ was observed.  
Interestingly, in canonical ferroelectrics the relaxation dynamics corresponding to such transition are typically observed in the GHz frequency range, indicating that relaxing dipolar entities are relatively large and slowly fluctuating microscopic regions spanning several unit cells. 
%\textcolor{black}{This observation confirms that short range dipolar order rather than long range FE order exists.} 
 %These microregions can be viewed as regions with ferroelectric polarizations but with slightly different Curie temperature. These can be also be compared with the polar nanoregions in relaxors. 
%We therefore conclude that in our material local ordering of dipoles occurs resulting in formation of superparaelectric dipoles that extend over several unit cells and can be compared with polar nanoregions in relaxors. 

%In conclusion, we have successfully grown orthorhombically distorted perovskite fluoride NaMnF$_3$ thin films on (100) SrTiO$_3$ single crystal substrates via MBE. 
%The films were single-phase with four in-plane and two out-of-plane twin domains. In-plane there were two additional twinned domains rotated by 1.4$^{\circ}$ in-plane with respect to one another. Magnetizaton data indicated the presence of weak ferromagnetic ordering with $T_N= 66$~K. 
%The frequency dispersion of the dielectric permittivity revealed a Debye-like relaxation with a broad distribution of relaxation times, while the temperature dependence of the characteristic relaxation times indicated an order-disorder phase transition occurring at 8~K. 
\textcolor{black}{Large dielectric dipolar regions could form because of the diffusion of a FE soft mode\cite{garcia2014geometric, 
 PhysRevLett.116.117202} as a result of the disorder in the NMF/SRO film.  Although the NMF/STO film has only a small strain, local measurements like piezoelectric force microscopy (PFM) need to be performed to verify ferroelectricity because of the lack of an electrically conductive bottom layer.
 In adddtion, further research is required to obtain a tetragonal $a_\textrm{o} = c_\textrm{o}$ lattice  distortion in NaMnF$_3$ thin films to verify the large ferroelectric polarization predicted by theory.\cite{garcia2014geometric,PhysRevLett.116.117202}}     
 
%primarily due to 
%formation of polar microscopic regions and this can be viewed as microregions each with slightly different Curie temperature.the 
%ordering of superparaelectric dipoles that extend over several unit cells.  
\textcolor{black}{See supplementary material for details of the growth procedure and structural characteristics of the films.}

%\begin{acknowledgments}
This work was supported by the National Science Foundation (grant \# 1434897) and the WVU Shared Research Facilities.  
We thank A.\ H.\ Romero and C.\ Chen for helpful discussions.
%\end{acknowledgments}

\bibliography{NMF_v12_8b}
\end{document}